\documentclass[10pt,letterpaper]{article}
\usepackage[top=0.85in,left=2.75in,footskip=0.75in,marginparwidth=2in]{geometry}

\usepackage[utf8]{inputenc}

\usepackage{cite}

\usepackage{nameref,hyperref}
\usepackage[numbers]{natbib}

\usepackage{microtype}
\DisableLigatures[f]{encoding = *, family = * }

\raggedright
\usepackage{changepage}

\usepackage[aboveskip=1pt,labelfont=bf,labelsep=period,singlelinecheck=off]{caption}

\makeatletter
\renewcommand{\@biblabel}[1]{\quad#1.}
\makeatother

\usepackage{lastpage,fancyhdr,graphicx}
\usepackage{epstopdf}
\fancyheadoffset[L]{2.25in}
\fancyfootoffset[L]{2.25in}

\usepackage{color}

\definecolor{Gray}{gray}{.25}

\usepackage{graphicx}

\usepackage{sidecap}
\usepackage{caption}
\usepackage{subcaption}

\usepackage{wrapfig}
\usepackage[pscoord]{eso-pic}
\usepackage[fulladjust]{marginnote}
\reversemarginpar
\usepackage{times}
\usepackage{wrapfig, frame}
\usepackage{multirow}

\begin{document}
\vspace*{0.35in}

% title goes here:
\begin{flushleft}
{\Large
\textbf\newline{Implementing Competency-Based Grading Improves the Performance of Women and First Generation Students in Introductory Physics}
}
\newline
% authors go here:
\\
Matthew Richard,
Jennifer Delgado,
Sarah LeGresley,
Christopher J. Fischer\textsuperscript{*}
\\
\bigskip
Department of Physics and Astronomy, University of Kansas, 1251 Wescoe Hall Drive, Lawrence, KS, 66049
\\
\bigskip
* shark@ku.edu

\end{flushleft}

\section*{Abstract}
    We present a model for competency-based grading for calculus-based introductory physics that encourages students to obtain proficiency with all course content. By allowing students to continually improve their proficiency with skills and content throughout the semester, this formative grading system is designed to create a more flexible learning environment that better accommodates the varying schedules and needs of students. While all students show improvement in their performance following the implementation of this grading system, the largest gains were found for women and first generation students, both of whom often pose a retention risk in science and engineering degree programs.

% now start line numbers
%\linenumbers

% the * after section prevents numbering
\section{Introduction}

The goal of this project was to develop a grading system for introductory physics that assesses student proficiency in different content areas of the course conditional on two criteria. First, students must demonstrate proficiency with content on more than one assessment. This stipulation is motivated by a desire to evaluate retention of knowledge and/or skills. Second, students who fail an initial assessment have sufficient subsequent opportunities to demonstrate their proficiency. This requirement stems from our experience with students achieving proficiency through different paths and/or at different rates owing to dissimilar prior experiences with course content, varying demands on their schedules, and other factors.  We found we could best satisfy both of these criteria by incorporating elements from competency-based grading models \cite{bloom1981all,colby1999grading,goubeaud2010science,dueck2011broke,shippy2013teaching,twyman2014competency,dueck2014grading} and mastery learning models \cite{carroll1963model,block1971mastery,block1974mastery,anderson1977mastery}. We report here our assessment of implementing this grading system in the first course of the calculus-based introductory physics sequence at the University of Kansas (KU).  While all students benefited from the switch to the new grading system, women and first generation students showed the largest gains in performance.

Further separate investigation of these data also allowed us to quantify the relative difficulty of each content area in PHSX 211.  We found that students struggle with force-based mechanics more than energy-based mechanics \citep{PhysRevPhysEducRes.15.020126}, and that quantitative problem solving with momentum presents the largest difficulty.  Taken together, these results are consistent with previously published studies \citep{knight1995vector,nguyen2003initial,Ozimek2005,mikula2013student,southey2014vector} and also suggest how instruction methods and time might be adjusted to better serve the needs of students.  Finally, these results are also another example of how compentency-based grading could be used in a course prerequisite system.

\section{Background and Motivation}

Several factors can influence the ability of a student to achieve proficiency in introductory physics including prior experience with mathematics and physics \cite{brekke1994some,hudson1977correlation,hudson1981correlation,Tyson2007}, stereotype threat \cite{betz2012my,madsen2013gender,marchand2013stereotype}, overall stress and mental load, etc. There also additional external pressures that can affect a student’s path to course proficiency. For example, an ever larger number of college and university students have significant demands on their time and attention outside of academics, such as working (full- or part-time), managing child- or adult-care, coping with mental illness or feelings of isolation, or struggling with food insecurity \cite{cook2010life,perna2010understanding,markle2015factors,conley2017meta,goldrick2017hungry,varghese2017college}.  These pressures and responsibilities can overwhelm students and prevent them from maintaining a consistent schedule for studying, completing homework assignments on time, or even attending class regularly. When taken together, these factors can result in a distribution of pathways and associated rates taken by students to achieve proficiency in the different content areas of their courses. In analogy to Carroll’s model of mastery-based instruction \cite{carroll1963model}, we would therefore expect that scores on assessments of student learning will likely follow a normal distribution if all students are required to demonstrate proficiency with course content on a fixed schedule of assessments. In contrast, we suspect, as Carroll argued, that if the schedule and/or number of assessments is adapted to the needs of the learner, the distribution of scores on these assessments should be skewed toward the high end of the achievement measure \cite{carroll1963model}.  Indeed, the predictions of Carroll’s model \cite{carroll1963model} are consistent with the results of subsequent studies in in K-12 classrooms \cite{anderson1977mastery,block1971mastery,block1974mastery}, including that the implementation of mastery-based instruction resulted in aptitude tests no longer having predictive power for student outcomes in a course.

While competency-based learning or elements thereof have been incorporated into university instruction in several disciplines \cite{cohen2016mentorship,harden1999amee,palardy1972competency,scott2008competency,voorhees2001competency,zeichner1983alternative}, there are limited reports of its application in the physical sciences, and we are aware of no published analysis of its use in calculus-based physics instruction. There is a previous study \cite{rajapaksha2017competency} of competency-based instruction in algebra-based introductory physics at Purdue University, however. In that study, a determination of student preparedness for using the mathematics associated with the course was obtained using a student survey. This survey, administered following initial instruction in the mathematical concepts and skills necessary for the course, assessed previous experience with this content and confidence in applying it. The authors of that study report that those students professing novice ability in or limited prior experience with the mathematics needed for introductory physics (i.e., those whom the authors deemed the least prepared to learn college physics) showed the largest gains in learning outcomes from the implementation of a competency-based grading system in introductory physics \cite{rajapaksha2017competency}.

We therefore posited that increasing the flexibility by which students demonstrate their proficiency with course content would result in improved student proficiency and overall performance in calculus-based introductory physics.  For this reason we developed a new grading system that afforded students multiple opportunities to display their competency with course material, rather than relying on more traditional single (midterm exam) or at best two (midterm exam and final exam) assessments of ability. We assessed the impact of this grading system on student performance through associated changes in course grade distributions and in the ``D'', fail, and withdraw (DFW) rate. In order to determine whether this grading system affected student populations differently, we further parsed these data into the following cohorts of students: men, women, URM students in physics (defined here to be Black, Hispanic, and Native American students) and first generation students.  It is important to note that since we are relying upon institutional data for student gender and ethnicity, these divisions do not therefore necessarily reflect how students identify personally.  Furthermore, a student declaring mixed ethnicity who included Black, Hispanic, and/or Native American on their subsequent list of all of their ethnic groups was considered a URM student in this study. 

\section{Grading Model}

The population enrolled in the introductory physics course in this study, General Physics I (PHSX 211), consists mostly of students majoring in the physical sciences and engineering. In the grading system for this course, a student’s grade is determined from the student’s achievement of competency (i.e., demonstration of proficiency) in distinct content areas for the course; we have named this grading system \textbf{A}ssessment of \textbf{L}earning \textbf{P}roficiency \textbf{a}nd \textbf{C}ompetency \textbf{A}chievement (ALPaCA). We use the following six content areas for PHSX 211: (1) kinematics (linear and rotational, constant and non-constant acceleration); (2) energy, energy conservation, and energy-based mechanics; (3) oscillatory motion; (4) Newton’s laws (forces and force-based mechanics); (5) momentum conservation (linear and rotational); and (6) thermodynamics.  We use broadly defined content areas to limit the number of content areas in the grading system and to more clearly indicate to students the key topics of importance to the course. The numbering of these content areas follows the chronological order of presentation of the associated topic during the semester \cite{PhysRevPhysEducRes.15.020126}.

\subsection{Initial Version of ALPaCA}

In the implementation of ALPaCA grading used in PHSX 211 during the collection of the data presented here, there were three section exams and one comprehensive final exam; students were allotted 75 minutes for each section exam and 150 minutes for the final exam. In order to afford students additional opportunities to demonstrate competency in course content, all exams were cumulative, as shown in Table \ref{tab:tab1}, and included questions for each content area covered in the course up to the time of the exam.  The total number of questions for each Content Area and corresponding number of total possible points available are also shown.

\begin{table}[htbp]
  \caption{The number of questions from each Content Area on each exam.  The total number of questions for each Content Area and corresponding number of total possible points available are also shown.\label{tab:tab1}}
    \begin{tabular}{c c c c c c c}
         \multirow{2}{*}{\textbf{Content Area}} & \multicolumn{5}{c}{\textbf{Number of Questions}} & \multirow{2}{*}{\textbf{Total Points}} \\
         & \textbf{Exam 1} & \textbf{Exam 2} & \textbf{Exam 3} & \textbf{Final} & \textbf{Total} & \\
         \hline
         \hline
         1 & 4 & 2 & 2 & 3 & 11 & 33\\
         2 & 4 & 2 & 2 & 3 & 11 & 33\\
         3 &  & 4 & 2 & 3 & 9 & 27 \\
         4 &  & 4 & 2 & 3 & 9 & 27 \\
         5 &  &   & 4 & 3 & 7 & 21 \\
         6 &  &   & 4 & 3 & 7 & 21 \\
    \end{tabular}
\end{table}

Each of the three section exams included an individual portion and a group portion, which were administered sequentially within the same class period. The same questions were used for both the individual and group portions of the exam so that, following the individual portion of the exam, students could discuss their answers as a group to reach consensus. The final exam had only an individual portion. Each correct answer on an exam earned points toward demonstrating competency with the content area associated with that question. A student earned three points for each correct answer on the individual portions of the exams. On the subsequent group potions of the exams, a student earned one point for each question answered correctly that was answered incorrectly on that student’s individual exam. For example, a student answering three of four questions for a single content area correctly on the individual portion of an exam will earn nine points. If that student then correctly answers all four questions for that content area on the group portion of the exam, then that student will earn an additional one point for that content area, for a total of ten points earned for that content area. The number of points accumulated for each content area is converted to a grade for each content area using the matrix in Table \ref{tab:tab2}.

\begin{table}[htbp]
  \caption{Calculation of student grades in PHSX 211 for each grading category of the course.  The minimum number of points in each content area for each grade is indicated.  We also assess student participation in the course both during class time (through their participation in course discussions and student-centered activities) and outside of class time (through their completion of homework).\label{tab:tab2}}
    \begin{tabular}{c c c c c c c}
         \multirow{2}{*}{\textbf{Category}} & \multirow{2}{*}{\textbf{Index (i)}} & \multicolumn{5}{c}{\textbf{Grade $\left( s_i \right)$}}  \\
         &  & 5 (A) & 4 (B) & 3 (C) & 2 (D) & 1 (F) \\
         \hline
         \hline
         Content Area \#1 & 1 & 20 & 17 & 14 & 11 & $< 11$\\
         Content Area \#2 & 2 & 20 & 17 & 14 & 11 & $< 11$\\
         Content Area \#3 & 3 & 15 & 12 & 9  & 6 & $< 6$ \\
         Content Area \#4 & 4 & 15 & 12 & 9  & 6 & $< 6$ \\
         Content Area \#5 & 5 & 10 & 8  & 6  & 4 & $< 4$ \\
         Content Area \#6 & 6 & 10 & 8  & 6  & 4 & $< 4$ \\
         Homework Score (\%) & 7 & $>74$  & 60 - 74 & 45 - 59 & 30 - 44 & $<30$ \\
         Class Participation (\%) & 8 & $>74$  & 60 - 74 & 45 - 59 & 30 - 44 & $<30$ \\
    \end{tabular}
\end{table}

As shown in Table \ref{tab:tab2}, we also included grades for class participation and homework completion. We assigned the point values for Content Area \#1 through Content Area \#4 in Table \ref{tab:tab2} using the following two criteria: (i) a student cannot obtain the points required for maximum grade of five (A) from a single exam and (ii) a student can obtain the points required for the maximum grade of five (A) without taking the final exam even if the student answered all questions correctly on the first section exam for that content area. The first criterion requires students to demonstrate their proficiency with a content area on more than one exam; this criterion thus assesses the retention of course content. The second criterion affords students additional opportunities to demonstrate proficiency with a content area if they are unable to do so on their first attempt; this criterion thus accommodates different rates of or schedules available for learning. These criteria cannot both be applied to Content Area \#5 and Content Area \#6, however, since these objectives are covered at the end of the semester. Thus, for these two content areas, a student can obtain the points required for the maximum grade of five by answering correctly all questions on the third section exam.

While our long term goal would be to replace an overall course grade with a list of separate grades for each content area, in the short term the university still requires us to provide a single grade for each student in the course. We have decided to calculate this final course grade from the geometric mean of the eight individual grades in Table \ref{tab:tab2} using Equation \ref{eq:signal} and Table \ref{tab:tab3}.  Thus, a student with scores $s_1=5$, $s_2=3$, $s_3=4$, $s_4=5$, $s_5=2$, $s_6=3$, $s_7=5$, $s_8=4$ would have $s_{final}=3.7$ and would earn a ``C'' in the course.

\begin{equation}
    s_{final} = \left( s_1 \cdot s_2 \cdot s_3 \cdot s_4 \cdot s_5 \cdot s_6 \cdot s_7 \cdot s_8 \right)^{\frac{1}{8}}
\end{equation}\label{eq:signal}

\begin{table}[htbp]
  \caption{Converting a final course score into a letter grade.\label{tab:tab3}}
    \begin{tabular}{c c}
        \textbf{Score} & \textbf{Letter Grade} \\
         \hline
         \hline
         $s_{final} \ge 4.6$ & A \\
         $3.9 \le s_{final} < 4.6$ & B \\
         $2.6 \le s_{final} < 3.9$ & C \\
         $1.6 \le s_{final} < 206$ & D \\
         $s_{final} < 1.6$ & F \\
    \end{tabular}
\end{table}

Lower proficiency scores negatively affect the course grade more under a geometric mean than under an arithmetic mean.  Because of this, calculating the course grade from the geometric mean results in students receiving a larger increase to their course grade from raising their lowest competency scores rather than their highest competency scores.  Indeed, even for changes of similar magnitude (i.e., shifting a score of 4 to a 5 versus shifting a score of 3 to a 4), increasing a lower competency score raises the course grade more than increasing a higher competency score.  Thus, this approach to calculating the course grade incentivizes students to continually improve their competency in all course content.  We therefore also hoped that this grading system would be formative for students by providing them clear feedback about what they need to do to improve their course grade. Finally, we hoped that switching from a curve-based grading system, that can encourage competition between students, to a competency-based grading system would instead encourage cooperation between students \cite{bloom1981all}, which should increase the usefulness of the group portions of the examinations and other in-class group activities.

\section{Results}

\begin{figure*}
  \includegraphics[width=0.8\textwidth]{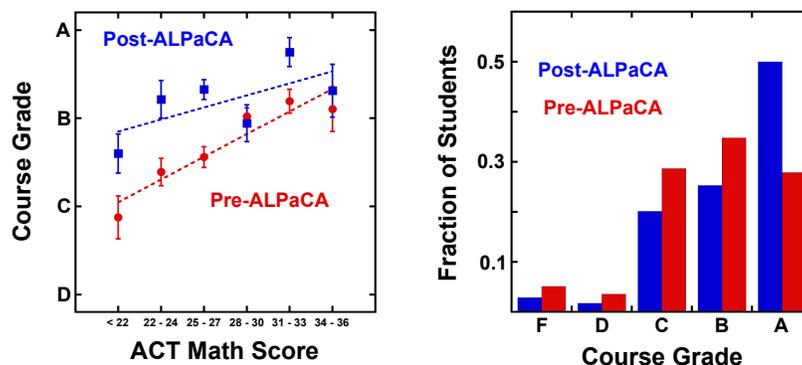}
  \caption{Average grades earned in PHSX 211, sorted by ACT math score, before (red) and after (blue) the implementation of ALPaCA grading (left panel). The dashed lines in this figure are linear fits of this data meant only to guide the eye. The fraction of students earning each course grade before (red) and after (blue) the implementation of ALPaCA grading (right panel).} \label{fig:fig1}
\end{figure*}

Since we implemented a new curriculum for PHSX 211 in the Spring 2015 semester \citep{PhysRevPhysEducRes.15.020126}, we analyzed the performance of only students completing PHSX 211 between the Spring 2015 semester and the Spring 2019 semester; the total course enrollment for these semesters is 727 students.  These data include seven semesters in which ALPaCA grading was not used (635 students between Spring 2015 and Spring 2018) and two semesters in which it was used (182 students in Fall 2018 and Spring 2019).  In addition to not associating exam questions with specific Content Areas, the pre-ALPaCA exams in PHSX 211 were not comprehensive.  Rather, they contained questions covering only a particular portion of the course content; the final exam was still comprehensive, however.  Thus, students had only one opportunity, outside of the comprehensive final, to demonstrate their proficiency with each element of course content.

\subsection{Effect on Grade Distribution}

As shown in Figure \ref{fig:fig1}, a positive correlation exists between the grade earned in PHSX 211 and student ACT math score, consistent with what has been observed in physics courses at other institutions \cite{brekke1994some,cohen1978cognitive,hudson1977correlation,hudson1981correlation}.  The data shown in Figure \ref{fig:fig1} also indicate that the implementation of ALPaCA grading weakened the dependence of course grade on ACT math score.  Since ACT math score can serve as a proxy for mathematics ability \cite{PhysRevPhysEducRes.15.020126}, these data suggest that course grade is less dependent upon prior mathematics ability when ALPaCA grading is used.  In addition to implying that ALPaCA grading may allow students to more effectively confront obstacles to learning physics associated with poor mathematics preparation, this observation is also consistent with previously published results demonstrating that competency-based grading in algebra-based introductory physics had the largest benefit for students with lowest mathematics ability \cite{rajapaksha2017competency}.  

The data in Figure \ref{fig:fig1} also show that the implementation of ALPaCA grading has affected the distribution of grades earned by students in PHSX 211.  While approximately the same fraction of students earn grades of ``A'', ``B'', or ``C'' in these courses (95\% pre-ALPaCA and 97\%  post-ALPaCA), a higher fraction of those students earn a grade of A post-ALPaCA.  Furthermore, the fraction of students earning a grade of ``A'' or ``B'' increased from 68\% pre-ALPaCA to 75\% post-ALPaCA, and the fraction of students earning a grade of ``A'' increased from 30\% pre-ALPaCA to 50\% post-ALPaCA.

\subsection{Effects on Different Student Populations}

\begin{figure}
     \captionsetup[subfigure]{justification=centering}
     \centering
     \begin{subfigure}[b]{0.45\textwidth}
         \centering
         \includegraphics[width=\textwidth]{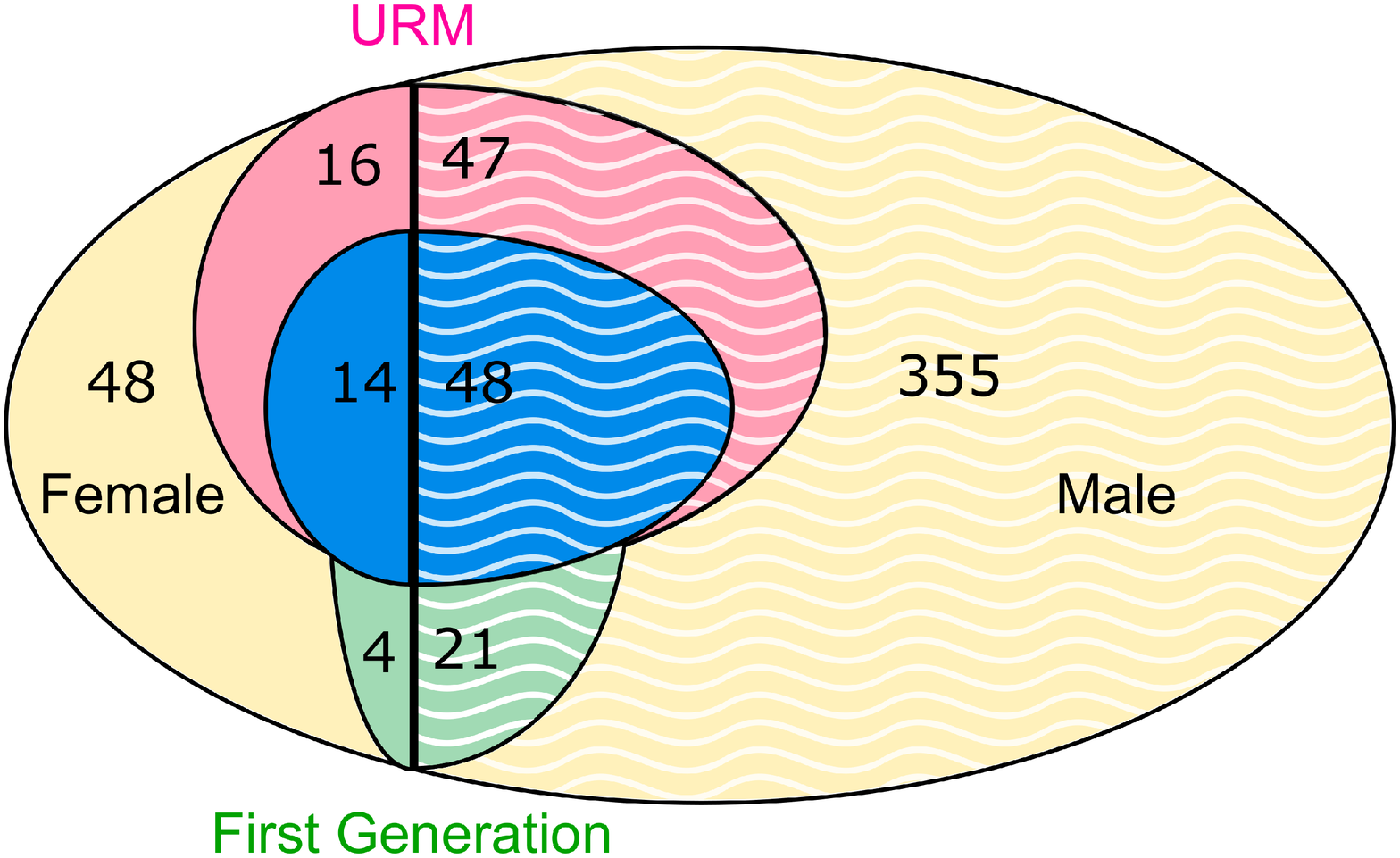}
         \caption{Pre-ALPaCA}
         \label{fig:fig2a}
     \end{subfigure}
     \hfill
     \begin{subfigure}[b]{0.45\textwidth}
         \centering
         \includegraphics[width=\textwidth]{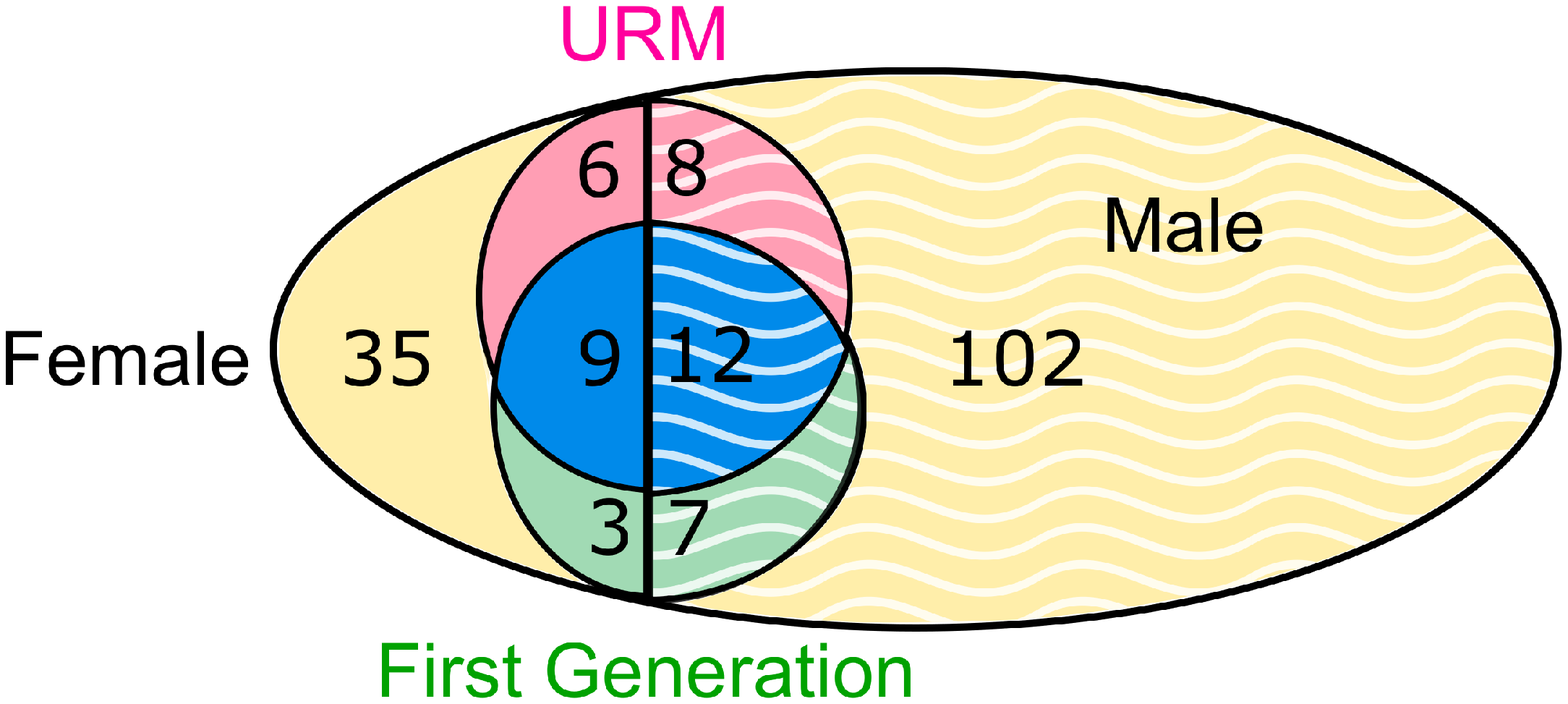}
         \caption{Post-ALPaCA}
         \label{fig:fig2b}
     \end{subfigure}
     \caption{Student enrollment numbers in PHSX 211.  The population of students who are URM, first generation, or URM and first generation are shaded pink, green, and blue, respectively.  Scale is not completely representative of the populations.}
     \label{fig:fig2}
\end{figure}

\begin{figure}
\begin{center}
\includegraphics[width=0.8\textwidth]{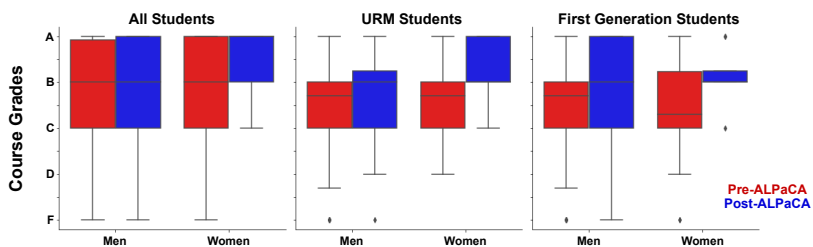}
\end{center}
\caption{ Average course grades earned in PHSX 211 for different cohorts of students (men, women, URM, and first generation) in PHSX 211 before (red) and after (blue) the implementation of ALPaCA grading.  Women and first generation students display more significant improvements in average course grade than their peers.}
\label{fig:fig3}
\end{figure}

We next determined the effect of ALPaCA grading on different cohorts of students: men, women, URM students, and first generation students; the distribution of these students is shown in Figure \ref{fig:fig2}.  We show in Figure \ref{fig:fig3} the average grades earned in PHSX 211 before and after the implementation of ALPaCA grading.  While these data indicate using ALPaCA grading resulted in an increase in the average course grade of each cohort of students, notable variation exists.  For example, the lower bound of grades earned by women (including women who were also URM or First Generation students) increased more than the lower bound of grades earned by men.  While both male and female first generation students experienced a boost in average course grade after the switch to ALPaCA grading, the course grades of female URM students see a larger benefit from ALPaCA grading than those of male URM students.

\begin{figure*}
  \centering
  \includegraphics[width=0.9\textwidth]{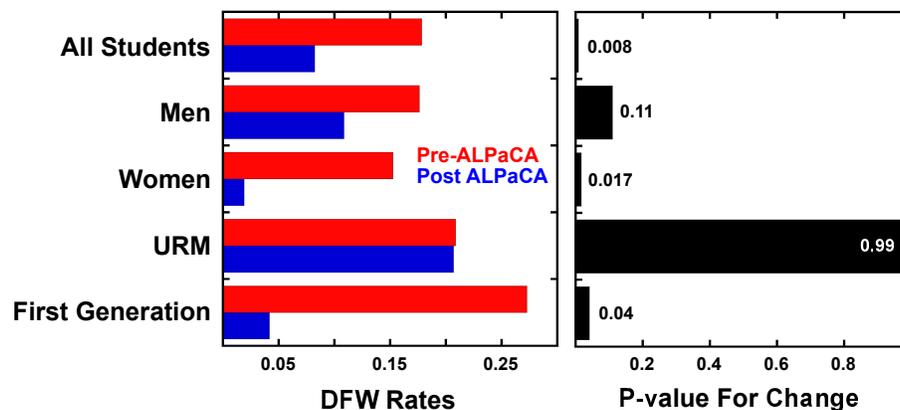}
  \caption{DFW rates for different cohorts of students in PHSX 211 before (red) and after (blue) the implementation of ALPaCA grading (left panel).  The p-values for the change in DFW rates following the implementation of ALPaCA grading (right panel). The most significant effects are seen for women and first generation students.}
  \label{fig:fig4}
\end{figure*}

We show in Figure \ref{fig:fig4} the DFW rates for each cohort of students before and after the implementation of ALPaCA grading as well as the p-values for the change in the rate following the switch to ALPaCA grading. While the DFW rate decreased for each cohort following the implementation of ALPaCA grading, the changes were most significant for women and first generation students.

\subsection{Assessing Student Performance across Course Content}

\begin{figure*}
  \centering
  \includegraphics[width=0.7\textwidth]{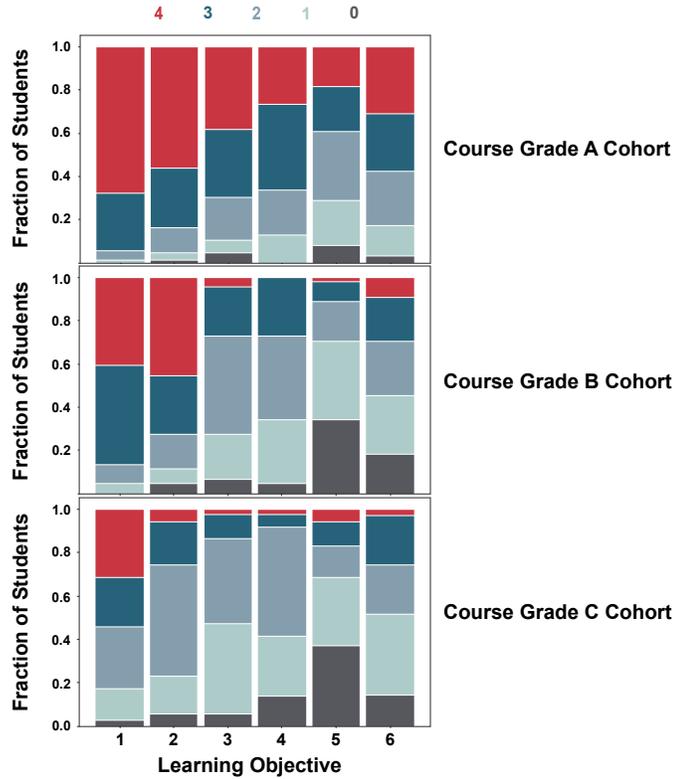}
  \caption{The distribution of correct answers for each content area when it is first assessed.  These data are further distributed into cohorts of students earning the same final course grade.} \label{fig:fig5}
\end{figure*}

\begin{figure*}
  \centering
  \includegraphics[width=0.6\textwidth]{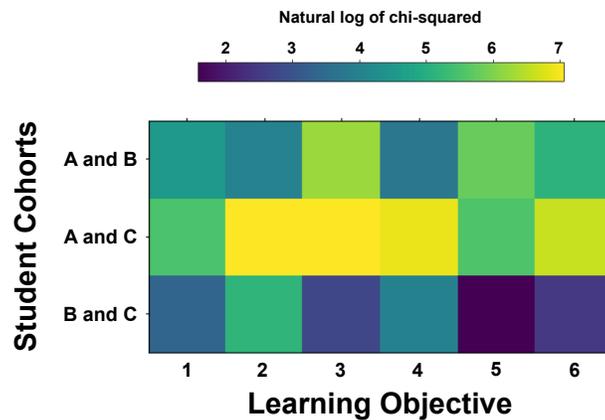}
  \caption{The natural log of the chi-squared values from comparisons of distributions of correct answers of different cohorts of students for different content areas. A lower value for chi-squared indicates more similar distributions.}
  \label{fig:fig6}
\end{figure*}

The first assessment of each content area consists of four exam questions as shown in Table \ref{tab:tab1}. We show in Figure \ref{fig:fig5} the distribution of students answering these questions correctly, further parsed according to the final course grade earned by the students; e.g., students in the Grade A Cohort earned a final grade of ``A'' in the course. These data  indicate that the difficulties of the content areas are non-uniform within and across these cohorts.   To investigate these differences further, we quantitatively compared the distributions of correct answers of different cohorts of students for different content areas by calculating the associated chi-squared $\left( \chi^2  \right)$.  The lower the $\chi^2$ the more similar the distributions, with a value of $\chi^2 = 0$ indicating that the distributions are identical.  For ease of presentation, we show the natural logarithm of these chi-squared values in Figure \ref{fig:fig6}.

This analysis indicates that Content Area \#1 (kinematics) shows the least discrimination between the cohorts. In other words, students perform similarly well on assessments of kinematics regardless of their final course grade. For Content Area \#4 (Newton’s laws), the difference in performance between the Grade A Cohort and Grade B Cohort is nearly identical to the difference in performance between the Grade B Cohort and the Grade C Cohort.  This linear behavior indicates that performance in this content area is well discriminated among the students.  In contrast, non-linear discrimination was found for Content Area \#3 (oscillatory motion), Content Area \#5 (momentum conservation), and Content Area \#6 (thermodynamics). These results were most stark for Content Area \#5, in which students in the Grade B Cohort and Grade C Cohort showed similar proficiency (84\% probability that these distributions were identical), but had significantly poorer performance than students in the Grade A Cohort. The similarity between the distributions for the Grade B Cohort and Grade C Cohort for Content Area \#3 and Content Area \#6 were smaller (21\% and 9\%, respectively), but in both cases these cohorts displayed significantly worse proficiency than the Grade A Cohort. The differences for both of these content areas were not as large as for Content Area \#5, however.

\subsection{Student Comments}

The University of Kansas uses student surveys as part of instructor and course evaluation.  In addition to multiple choice questions concerning various aspects of student perception of the course and instructor effectiveness, these surveys also contain free response sections for students to comment on aspects of the course they liked and changes they would suggest.  In the Fall 2018 semester, 56 students in PHSX 211 (73\% of total course enrollment) filled out the free response section of the survey.  Fourteen of those students commented on the use of objective-based grading in the course and all of those comments were positive.  Similarly, in the Spring 2019 semester, 51 students in PHSX 211 (49\% of total course enrollment) filled out the free response section of the survey.  Eighteen of those students commented on the use of objective-based grading in the course and all of those comments were positive.  The following comments are a selection of those appearing on these surveys:

\begin{quotation}
\noindent \textit{The grading system is much better than in other classes.}

\vspace{2mm}

\noindent \textit{I liked the grading system a lot more than in traditional courses.}

\vspace{2mm}

\noindent \textit{Objective-based grading works extremely well.}

\vspace{2mm}

\noindent \textit{I definitely learned more because I had to keep reviewing problems and concepts.}
\end{quotation}

\noindent Thus, although the structure of ALPaCA grading is different from traditional grading methodologies employed in most courses, students nevertheless have had a positive reaction to it.

\section{Discussion}

Although ALPaCA does not include all elements of mastery-based instruction, the results of ALPaCA implementation are nevertheless consistent with previous work on it. Specifically, the observed increase in the fraction of students earning a grade of ``A'' after the implementation of ALPaCA grading (Figure \ref{fig:fig1}) is consistent with the predictions of Carroll’s model \cite{carroll1963model}  and the observed weakening of the dependence of PHSX 211 course grade on ACT math score aligns well with previous studies linking mastery-based instruction with a decrease in the predictive power of aptitude tests \cite{anderson1977mastery,block1971mastery,block1974mastery}.  Furthermore, in context of Carroll’s model \cite{carroll1963model}, our observation that women and first generation students show substantial benefits from ALPaCA grading, suggests that the use of ALPaCA grading may be creating a more supportive/flexible environment for these cohorts of students.  In contrast, the less dramatic benefit observed for URM students in our study suggests that implementing ALPaCA grading has not as significantly addressed the dimensions of the course that negatively affect the performance of this cohort of students. However, that female URM students seem to benefit more than male URM students may indicate that different intersections of identities modulate how ALPaCA grading impacts students. It is also possible, of course, that some other aspect of the course more strongly influences the performance of URM students than ALPaCA grading. More data are clearly required to deconvolute further how various dimensions of the course affect student performance.

\subsection{Applications to Course Assessment}

In addition to potentially improving student performance, especially for some traditionally under-served populations, the implementation of ALPaCA grading has also improved our course assessment of PHSX 211.  As mentioned previously, we began using a new curriculum for PHSX 211 in the Spring 2015 semester \citep{PhysRevPhysEducRes.15.020126}.  This curriculum shifts the initial focus of instruction away from forces and the associated vector mathematics to the scalar quantity energy and associated differential and integral calculus \citep{Fischer2017}.  The data presented here (Figure \ref{fig:fig5}) demonstrate that students perform better solving classical mechanics problems that are energy-based (Content Area \#2) than those that are force-based (Content Area \#4), even though the students are required to use calculus more often when solving the energy-based problems than the force-based problems.  This result is consistent with numerous previous studies documenting student difficulty with the vector mathematics required to use forces correctly \citep{knight1995vector,nguyen2003initial,Ozimek2005,mikula2013student}.  Our observation that students struggle most solving problems involving linear momentum and angular momentum (Content Area \#5) is also consistent with the results of previous studies \citep{southey2014vector} and might result from these concepts and solutions requiring competency with both calculus and vectors \citep{PhysRevPhysEducRes.15.020126}.  We suggest that this dual requirement also explains why performance on Content Area \#5 provides such high discrimination between students earning a grade of ``A'' in PHSX 211 and students earning other grades (Figure \ref{fig:fig6}).  These data also suggest that students in PHSX 211 might be better served if less time was devoted to kinematics (Content Area \#1) and more time was devoted on linear and angular momentum (Content Area \#5).

The data presented in Figure \ref{fig:fig5} also suggest when interventions, such as supplemental instruction, might be most effective. For example, the first assessments of Content Area \#2 and Content Area \#3 occur early in course. Since performance on these assessments provides good discrimination between final course grades, poor performance on these initial evaluations of these content areas might serve as a trigger for the implementation of additional tutorials or other supplemental materials.

\subsection{Limitations}

While the results presented here demonstrate potential benefits of implementing ALPaCA grading in introductory physics, it is important to note the limitations of what we have done and the modifications we are planning to pursue in the future.  First, comparisons of student performance (grade distributions and DFW rates) before and after the implementation of ALPaCA grading are problematic due to the differences in grading methodologies used.  For example, under pre-ALPaCA grading students had only one opportunity, outside of the comprehensive final, to demonstrate their proficiency with each element of course content, whereas under ALPaCA grading students have multiple opportunities.  Furthermore, course grades were curved pre-ALPaCA, but not under ALPaCA grading.

Second, our choice of the point thresholds for each score for each content area and/or the distribution of questions on the exams may not be appropriate for assessing student proficiency with the content areas of the course.  Lastly, it's possible that these Content Areas may be too broadly defined to effectively assess student proficiency with course content or in some other way be inappropriate for our larger goal of assessing student proficiency as part of a prerequisite structure for downstream courses.

\section{Future Work}

We intend to explore further the mechanism(s) by which ALPaCA grading results in improvements in student performance and why the scales of these improvements vary among student cohorts. In the short term, we plan to use existing instruments \cite{marshman2018female,lindstrom2011self,shaw2004development,marshman2018longitudinal,beatty2020improving} to assess how student self-efficacy with physics \cite{fencl2004pedagogical,lindstrom2011self,sawtelle2012exploring,shaw2004development} and mindset \cite{cato2011mindset,mcclendon2017grit,aguilar2014psychological} are affected by the implementation of ALPaCA grading.  Since self-efficacy and mindset are both linked with student performance \cite{malespina2022whose,sawtelle2012exploring}, we hypothesize that students will experience a more positive change (or less negative change) in their physics self-efficacy when ALPaCA is used.  Similarly, we anticipate a larger fraction of the students will display a growth mindset when ALPaCA is used.  It will be of particular interest to determine whether the identity correlated changes in student performance following the implementation of ALPaCA grading would be associated with similar identity correlated changes in self-efficacy or mindset.

\subsection{Applications to Prerequisite Models}

Monitoring changes to self-efficacy and mindset will also be a component of additional future longitudinal studies in which we track student performance in course content areas throughout the curriculum of their degree plans.  This will also enable us to quantify the predictive power of student proficiency in one course's content areas on downstream performance in other courses, both inside and outside of physics. This will including determining not only how student proficiency with individual content areas in PHSX 211 affects downstream course grades, but also which combinations of content areas are most relevant for different subsequent courses.  This will also allow us to further refine the list of content areas that are most important/useful for PHSX 211 and what formula should be used to calculate a course grade from student proficiency with these content areas.  This is also the necessary next step in the development of a new prerequisite model for course scheduling and advising based on student proficiencies rather than course completions.

Finally, we plan to expand ALPaCA grading to other courses within our department so that we can monitor improvements in student competency with the content areas of our degree programs.  This will also provide a mechanism by which we can more readily link different courses in a sequence together and thereby better demonstrate to the students how the content of these courses map into the content areas of their major.  If successful, this can become a template for other departments to use in the assessment of student performance and development within their disciplines.

\subsection{Refinements to ALPaCA}

We have continued to refine the ALPaCA method since the implementation presented here is cumbersome and somewhat complicated.  In the latest iteration, we use shorter quizzes, administered weekly throughout the semester, rather than the longer exams described here.  Each quiz is focused on a single content area and at least two quizzes occur each week.  The score used for each content area in the calculation of the course grade using the geometric mean is the highest of the score earned on the quizzes (more specifically, the average of the two highest scores earned on that content area’s quizzes) or on the final exam.  Using the average of the two highest scores addresses the requirement that students demonstrate proficiency on more than one assessment and having multiple quizzes throughout the semester (including the final exam, which consists of one quiz per content area) affords students multiple opportunities to demonstrate proficiency.  All quizzes and the final exam are individual assessments as we have found limited benefit from group exams.  We retain group work, however, as part of the active-learning instruction in the course.  

This system allows students to demonstrate maximum proficiency (i.e., earn an ``A'' in the course) without having to take the final exam or by doing well on only the final exam.  Providing students multiple pathways to demonstrate proficiency reduces the stress associated with doing well on every assessment and focuses attention on what proficiency with course content students have demonstrated by the end of the semester.  This simplified version of ALPaCA has been better received by students as the method of calculating course grades is more direct and thus more readily understood.

%\clearpage

\section*{Acknowledgments}
We thank Professor Christopher Rogan for suggesting the use of a geometric mean rather than an arithmetic mean when calculating course grades.  We would also like to thank Adam Dubinksy and Associate Dean Holly Storkel for providing aggregated institutional data for this project.

%\nolinenumbers

%This is where your bibliography is generated. Make sure that your .bib file is actually called library.bib
\bibliography{library_ALPaCA}

\begin{thebibliography}{10}

\bibitem{aguilar2014psychological}
L.~Aguilar, G.~Walton, and C.~Wieman.
\newblock Psychological insights for improved physics teaching.
\newblock {\em Physics Today}, 67(5):43--49, 2014.

\bibitem{anderson1977mastery}
L.~W. Anderson and J.~H. Block.
\newblock Mastery learning.
\newblock In {\em Handbook on teaching educational psychology}, pages 163--185.
  Elsevier, 1977.

\bibitem{beatty2020improving}
I.~D. Beatty, S.~J. Sedberry, W.~J. Gerace, J.~E. Strickhouser, M.~A. Elobeid,
  and M.~J. Kane.
\newblock Improving stem self-efficacy with a scalable classroom intervention
  targeting growth mindset and success attribution.
\newblock In {\em Proceedings of the 2019 Physics Education Research
  Conference}, 2020.

\bibitem{betz2012my}
D.~E. Betz and D.~Sekaquaptewa.
\newblock My fair physicist? feminine math and science role models demotivate
  young girls.
\newblock {\em Social psychological and personality science}, 3(6):738--746,
  2012.

\bibitem{block1971mastery}
J.~H. Block and P.~W. Airasian.
\newblock {\em Mastery learning: Theory and practice}.
\newblock Holt Rinehart \& Winston, 1971.

\bibitem{block1974mastery}
J.~H. Block and L.~Anderson.
\newblock Mastery learning.
\newblock {\em Handbook on Teaching Educational Psychology}, 1974.

\bibitem{bloom1981all}
B.~S. Bloom.
\newblock {\em All our children learning: A primer for parents, teachers, and
  other educators}.
\newblock McGraw-Hill New York, 1981.

\bibitem{brekke1994some}
S.~E. Brekke.
\newblock Some factors affecting student performance in physics.
\newblock {\em Spectrum}, 01 1994.

\bibitem{carroll1963model}
J.~B. Carroll.
\newblock A model of school learning.
\newblock {\em Teachers college record}, 1963.

\bibitem{cato2011mindset}
J.~Cato.
\newblock Mindset matters.
\newblock {\em The Physics Teacher}, 49(1):60--60, 2011.

\bibitem{cohen1978cognitive}
H.~D. Cohen, D.~F. Hillman, and R.~M. Agne.
\newblock Cognitive level and college physics achievement.
\newblock {\em American Journal of Physics}, 46(10):1026--1029, 1978.

\bibitem{cohen2016mentorship}
J.~Cohen and A.~Kassam.
\newblock Mentorship for residents in psychiatry: a competency-based medical
  education perspective with career counseling tools.
\newblock {\em Academic Psychiatry}, 40(3):441--447, 2016.

\bibitem{colby1999grading}
S.~A. Colby.
\newblock Grading in a standards-based system.
\newblock {\em Educational Leadership}, 56(6):52--55, 1999.

\bibitem{conley2017meta}
C.~S. Conley, J.~B. Shapiro, A.~C. Kirsch, and J.~A. Durlak.
\newblock A meta-analysis of indicated mental health prevention programs for
  at-risk higher education students.
\newblock {\em Journal of counseling Psychology}, 64(2):121, 2017.

\bibitem{cook2010life}
S.~L. Cook and K.~Krupar.
\newblock Life ate my homework.
\newblock {\em Academe}, 2010.

\bibitem{dueck2011broke}
M.~Dueck.
\newblock How i broke my own rule and learned to give retests.
\newblock {\em Educational Leadership}, 69(3):72--75, 2011.

\bibitem{dueck2014grading}
M.~Dueck.
\newblock {\em Grading smarter, not harder: Assessment strategies that motivate
  kids and help them learn}.
\newblock ASCD, 2014.

\bibitem{fencl2004pedagogical}
H.~S. Fencl and K.~R. Scheel.
\newblock Pedagogical approaches, contextual variables, and the development of
  student self-efficacy in undergraduate physics courses.
\newblock In {\em AIP Conference Proceedings}, volume 720, pages 173--176.
  American Institute of Physics, 2004.

\bibitem{Fischer2017}
C.~J. Fischer.
\newblock {\em {The Energy of Physics, Part II: Electricity and Magnetism}}.
\newblock Cognella, Inc., 2017.

\bibitem{goldrick2017hungry}
S.~Goldrick-Rab, J.~Richardson, and A.~Hernandez.
\newblock Hungry and homeless in college: Results from a national study of
  basic needs insecurity in higher education.
\newblock 2017.

\bibitem{goubeaud2010science}
K.~Goubeaud.
\newblock How is science learning assessed at the postsecondary level?
  assessment and grading practices in college biology, chemistry and physics.
\newblock {\em Journal of Science Education and Technology}, 19(3):237--245,
  2010.

\bibitem{harden1999amee}
J.~Harden, M.~Crosby, M.~Davis, and R.~Friedman.
\newblock Amee guide no. 14: Outcome-based education: Part 5-from competency to
  meta-competency: a model for the specification of learning outcomes.
\newblock {\em Medical teacher}, 21(6):546--552, 1999.

\bibitem{hudson1977correlation}
H.~Hudson and W.~McIntire.
\newblock Correlation between mathematical skills and success in physics.
\newblock {\em American Journal of Physics}, 45(5):470--471, 1977.

\bibitem{hudson1981correlation}
H.~Hudson and R.~M. Rottmann.
\newblock Correlation between performance in physics and prior mathematics
  knowledge.
\newblock {\em Journal of Research in Science Teaching}, 18(4):291--294, 1981.

\bibitem{knight1995vector}
R.~D. Knight.
\newblock The vector knowledge of beginning physics students.
\newblock {\em The physics teacher}, 33(2):74--77, 1995.

\bibitem{PhysRevPhysEducRes.15.020126}
S.~E. LeGresley, J.~A. Delgado, C.~R. Bruner, M.~J. Murray, and C.~J. Fischer.
\newblock Calculus-enhanced energy-first curriculum for introductory physics
  improves student performance locally and in downstream courses.
\newblock {\em Phys. Rev. Phys. Educ. Res.}, 15:020126, Sep 2019.

\bibitem{lindstrom2011self}
C.~Lindstr{\o}m and M.~D. Sharma.
\newblock Self-efficacy of first year university physics students: Do gender
  and prior formal instruction in physics matter?
\newblock {\em International Journal of Innovation in Science and Mathematics
  Education}, 19(2), 2011.

\bibitem{madsen2013gender}
A.~Madsen, S.~B. McKagan, and E.~C. Sayre.
\newblock Gender gap on concept inventories in physics: What is consistent,
  what is inconsistent, and what factors influence the gap?
\newblock {\em Physical Review Special Topics-Physics Education Research},
  9(2):020121, 2013.

\bibitem{malespina2022whose}
A.~Malespina, C.~D. Schunn, and C.~Singh.
\newblock Whose ability and growth matter? gender, mindset and performance in
  physics.
\newblock {\em International Journal of STEM Education}, 9(1):1--16, 2022.

\bibitem{marchand2013stereotype}
G.~C. Marchand and G.~Taasoobshirazi.
\newblock Stereotype threat and women's performance in physics.
\newblock {\em International Journal of Science Education}, 35(18):3050--3061,
  2013.

\bibitem{markle2015factors}
G.~Markle.
\newblock Factors influencing persistence among nontraditional university
  students.
\newblock {\em Adult Education Quarterly}, 65(3):267--285, 2015.

\bibitem{marshman2018longitudinal}
E.~Marshman, Z.~Y. Kalender, C.~Schunn, T.~Nokes-Malach, and C.~Singh.
\newblock A longitudinal analysis of students’ motivational characteristics
  in introductory physics courses: Gender differences.
\newblock {\em Canadian Journal of Physics}, 96(4):391--405, 2018.

\bibitem{marshman2018female}
E.~M. Marshman, Z.~Y. Kalender, T.~Nokes-Malach, C.~Schunn, and C.~Singh.
\newblock Female students with a’s have similar physics self-efficacy as male
  students with c’s in introductory courses: A cause for alarm?
\newblock {\em Physical Review Physics Education Research}, 14(2):020123, 2018.

\bibitem{mcclendon2017grit}
C.~McClendon, R.~M. Neugebauer, and A.~King.
\newblock Grit, growth mindset, and deliberate practice in online learning.
\newblock {\em Journal of Instructional Research}, 8:8--17, 2017.

\bibitem{mikula2013student}
B.~D. Mikula and A.~F. Heckler.
\newblock Student difficulties with trigonometric vector components persist in
  multiple populations.
\newblock In {\em Proceedings of the 2013 Physics Education Research
  Conference, Portland, OR}, pages 253--256, 2013.

\bibitem{nguyen2003initial}
N.-L. Nguyen and D.~E. Meltzer.
\newblock Initial understanding of vector concepts among students in
  introductory physics courses.
\newblock {\em American Journal of Physics}, 71(6):630--638, 2003.

\bibitem{Ozimek2005}
D.~J. Ozimek.
\newblock {Retention and Transfer from Trigonometry to Physics}.
\newblock In {\em AIP Conference Proceedings}, volume 790, pages 173--176. AIP,
  2005.

\bibitem{palardy1972competency}
J.~M. Palardy and J.~E. Eisele.
\newblock Competency based education.
\newblock {\em The Clearing House: A Journal of Educational Strategies, Issues
  and Ideas}, 46(9):545--548, 1972.

\bibitem{perna2010understanding}
L.~W. Perna.
\newblock Understanding the working college student.
\newblock {\em Academe}, 96(4):30--33, 2010.

\bibitem{rajapaksha2017competency}
A.~Rajapaksha and A.~S. Hirsch.
\newblock Competency based teaching of college physics: The philosophy and the
  practice.
\newblock {\em Physical Review Physics Education Research}, 13(2):020130, 2017.

\bibitem{sawtelle2012exploring}
V.~Sawtelle, E.~Brewe, and L.~H. Kramer.
\newblock Exploring the relationship between self-efficacy and retention in
  introductory physics.
\newblock {\em Journal of research in science teaching}, 49(9):1096--1121,
  2012.

\bibitem{scott2008competency}
D.~D. Scott~Tilley.
\newblock Competency in nursing: A concept analysis.
\newblock {\em The journal of continuing education in nursing}, 39(2):58--64,
  2008.

\bibitem{shaw2004development}
K.~A. Shaw.
\newblock The development of a physics self-efficacy instrument for use in the
  introductory classroom.
\newblock In {\em AIP Conference Proceedings}, volume 720, pages 137--140.
  American Institute of Physics, 2004.

\bibitem{shippy2013teaching}
N.~Shippy, B.~A. Washer, and B.~Perrin.
\newblock Teaching with the end in mind: The role of standards-based grading.
\newblock {\em Journal of Family \& Consumer Sciences}, 105(2):14--16, 2013.

\bibitem{southey2014vector}
P.~Southey and S.~Allie.
\newblock Vector addition in different contexts.
\newblock In {\em 2014 Physics Education Research Conference Proceedings},
  pages 243--246, 2014.

\bibitem{twyman2014competency}
J.~S. Twyman.
\newblock Competency-based education: Supporting personalized learning.
  connect: Making learning personal.
\newblock {\em Center on Innovations in Learning, Temple University}, 2014.

\bibitem{Tyson2007}
W.~Tyson, R.~Lee, K.~M. Borman, and M.~A. Hanson.
\newblock {Science, Technology, Engineering, and Mathematics (STEM) Pathways:
  High School Science and Math Coursework and Postsecondary Degree Attainment}.
\newblock {\em Journal of Education for Students Placed at Risk (JESPAR)},
  12(3):243--270, oct 2007.

\bibitem{varghese2017college}
M.~E. Varghese and M.~C. Pistole.
\newblock College student cyberbullying: Self-esteem, depression, loneliness,
  and attachment.
\newblock {\em Journal of College Counseling}, 20(1):7--21, 2017.

\bibitem{voorhees2001competency}
R.~A. Voorhees.
\newblock Competency-based learning models: A necessary future.
\newblock {\em New directions for institutional research}, 2001(110):5--13,
  2001.

\bibitem{zeichner1983alternative}
K.~M. Zeichner.
\newblock Alternative paradigms of teacher education.
\newblock {\em Journal of teacher education}, 34(3):3--9, 1983.

\end{thebibliography}

%This defines the bibliographies style. Search online for a list of available styles.
\bibliographystyle{abbrv}

\end{document}